\documentclass{article}
\usepackage{times}
\usepackage{amsfonts}
\usepackage{graphicx}
\usepackage[pdfmark]{hyperref}
\begin{document}
\noindent
{\Large  WIGNER FUNCTIONS, FRESNEL OPTICS, AND\\ SYMPLECTIC CONNECTIONS ON PHASE SPACE}
\vskip1cm
\noindent
{\bf Jos\'e M. Isidro}\\
Grupo de Modelizaci\'on Interdisciplinar, Instituto de Matem\'atica Pura y Aplicada,\\ Universidad Polit\'ecnica de Valencia, Valencia 46022, Spain\\
{\tt joissan@mat.upv.es}
%\vskip1cm
%\noindent
%\today
\vskip1cm
\noindent
{\bf Abstract} We prove that Wigner functions contain a symplectic connection. The latter covariantises the symplectic exterior derivative on phase space. We analyse the role played by this connection and introduce the notion of {\it local symplectic covariance of quantum--mechanical states}. This latter symmetry is at work in the Schroedinger equation on phase space.

%\vskip1cm
%\tableofcontents

\section{Introduction}\label{bernabeudemierdeu}

A celebrated theorem by Stone and von Neumann states that every unitary, irreducible representation of the Heisenberg algebra 
\begin{equation}
[Q,P]={\rm i}\hbar
\label{bernabeumekagueuentuputamadreu}
\end{equation}
is unitarily equivalent to that in which $Q$ acts by multiplication and $P$ by differentiation of square--integrable, $q$--dependent wavefunctions $\psi$ \cite{NEUMANN}. This fact, supplemented with the uncertainty principle, surely discouraged physicists from considering phase--space formulations of quantum mechanics. Wigner stands out among those not discouraged. In his study of quantum corrections to classical statistical mechanics  \cite{WIGNER}, Wigner defined a function on classical phase space,
\begin{equation}
W_{\psi}(q,p):=\frac{1}{2\pi\hbar}\int{\rm d}y\,\psi^*\left(q-\frac{1}{2}y\right)\psi\left(q+\frac{1}{2}y\right){\rm e}^{-{\rm i}py/\hbar},
\label{bernabeumarikoneu}
\end{equation}
that enjoys properties similar to those of probability distributions. Although little known, this formulation of quantum mechanics on phase space has a considerable theoretical interest \cite{WIGNERKIM, ZACHOS} as well as useful practical applications \cite{SCHLEICH}. In general the integral (\ref{bernabeumarikoneu}) is difficult to compute, if altogether possible. However, assume that the wavefunction $\psi$ can be factorised as
\begin{equation}
\psi(q)={\rm e}^{-aq^2}\phi(q), 
\label{bernabeuketefostieu}
\end{equation}
with $a>0$ a dimensionful constant and $\phi(q)$ piecewise smooth, such that
\begin{equation}
\int_{-\infty}^{\infty}{\rm d}q\,{\rm e}^{-2aq^2}\phi^2(q)<\infty.
\label{bernabeumafioseudemierdeu}
\end{equation}
{}For example, the harmonic oscillator and the Gaussian wavepacket satisfy conditions (\ref{bernabeuketefostieu}) and (\ref{bernabeumafioseudemierdeu}). Then, expanding the wavefunction $\psi$ in a basis of oscillator eigenstates, Wigner's $W_{\psi}(q,p)$ can be recast in an integral--free form as \cite{TEGMEN}
\begin{equation}
W_{\psi}(q,p)=\frac{1}{\hbar\sqrt{2\pi a}}\,{\rm e}^{-2aq^2}\phi^*\left(q-\frac{{\rm i}\hbar}{2}{\partial_p}\right)\phi\left(q+\frac{{\rm i}\hbar}{2}{\partial_p}\right){\rm e}^{-p^2/2a\hbar^2}.
\label{bernabeumekagoentuputasombrahijoputa}
\end{equation}
The operator within the argument of $\phi$ can be canonically transformed into $q/2+{\rm i}\hbar\partial_p$, a convention more useful for our purposes. Then the two operators
\begin{equation}
Q_{A'_0}:=\frac{q}{2}+{\rm i}\hbar\partial_p, \qquad P_{A'_0}:=\frac{p}{2}-{\rm i}\hbar\partial_q
\label{losdebernabeusoisunapanda}
\end{equation}
also satisfy the Heisenberg algebra (\ref{bernabeumekagueuentuputamadreu}).

Let us for the moment pretend that we are unaware of the Stone--von Neumann theorem; let us also not be discouraged by the fact that $Q_{A'_0}$ and $P_{A'_0}$ are somewhat more complicated, in their action on wavefunctions, than the usual $Q\psi(q)=q\psi(q)$ and $P\psi(q)=-{\rm i}\hbar\partial_q\psi(q)$. This action will in fact require  wavefunctions $\Psi(q,p)$ that depend both on $q$ and $p$; we will see presently how to interpret them. On phase space we have the usual exterior derivative
\begin{equation}
{\rm d}:={\rm d}q\partial_q+{\rm d}p\partial_p
\label{pandademafiososinutiles}
\end{equation}
and the {\it symplectic}\/ exterior derivative
\begin{equation}
{\rm d}':=-{\rm d}q\partial_q+{\rm d}p\partial_p.
\label{chupapollasdemierda}
\end{equation}
The negative sign before the first term above is ultimately related to the antisymmetry of the quantum commutator (\ref{bernabeumekagueuentuputamadreu}) under the exchange of position and momentum or, equivalently, to the antisymmetry of the classical symplectic form. Consider the following connection 1--form on phase space:
\begin{equation}
A'_0:=\frac{1}{2{\rm i}\hbar}\left(p{\rm d}q+q{\rm d}p\right).
\label{ketefolleubernabeu}
\end{equation}
The operators (\ref{losdebernabeusoisunapanda}) are the result of covariantising the symplectic derivative ${\rm d}'$ by the connection $A_0'$:
\begin{equation}
{\rm i}\hbar D'_{A_0}:={\rm d}q\left(\frac{p}{2}-{\rm i}\hbar\partial_q\right)+{\rm d}p\left(\frac{q}{2}+{\rm i}\hbar\partial_p\right).
\label{bernabeuhundeteenlamierdeu}
\end{equation}
We are thus drawn to the conclusion that Wigner's formulation of quantum mechanics on phase space leads naturally to a covariantisation of symplectic derivatives. In plain words, Wigner functions carry a symplectic connection hidden inside. \footnote{Since we are covariantising the symplectic exterior derivative (\ref{chupapollasdemierda}), will denote all related quantities with a prime. This will help avoid confusion with gauge theories on fibre bundles, where one covariantises the usual exterior derivative (\ref{pandademafiososinutiles}).}

Now the hypotheses (\ref{bernabeuketefostieu}) and (\ref{bernabeumafioseudemierdeu}) need not always be satisfied. It makes sense to assume that, if the integral (\ref{bernabeumarikoneu}) is to be computed in a situation more general than that corresponding to eqns. (\ref{bernabeuketefostieu}) and (\ref{bernabeumafioseudemierdeu}),
connections $A'$ more general than (\ref{ketefolleubernabeu}) must also be taken into account. We will therefore covariantise ${\rm d}'$ as per
\begin{equation}
{\rm d}'\rightarrow D'_{A'}:={\rm d}'+A',
\label{bernabeuchupameelboleu}
\end{equation}
where
\begin{equation}
A'=\frac{1}{{\rm i}\hbar}\left[A'_q(q,p){\rm d}q+A'_p(q,p){\rm d}p\right]
\label{ramallochupameelcarallo}
\end{equation}
is a certain symplectic connection. As a rule, connections and the corresponding covariant derivatives arise whenever a local gauge symmetry is present \cite{MS}. It is the purpose of this article to elucidate what this gauge symmetry is and how it acts on phase space.

Additional motivatation for our analysis comes from the following, apparently unrelated fact. It is well known that geometrical optics is to wave optics as classical mechanics is to quantum mechanics \cite{LANDAUQM}. Further pursuing this analogy, Wigner's approach to quantum mechanics on phase space has been argued \cite{FRESNEL} to be the analogue of Fresnel optics, {\it i.e.}, a wave theory of phenomena in which terms up to quadratic powers are taken into account, and higher powers are neglected. This quadratic truncation of power--series expansions in the relevant variables ($q$ and $p$ in the case of phase space) yields precisely the regime in which the semiclassical WKB approximation is actually exact. We observed above that the derivation of eqn. (\ref{bernabeumekagoentuputasombrahijoputa}) starting from the general Wigner integral (\ref{bernabeumarikoneu}) hinges crucially on the expansion of $\psi$ into oscillator eigenstates. It follows that eqn. (\ref{bernabeumekagoentuputasombrahijoputa}) can be regarded as a semiclassical expression of the general Wigner integral (\ref{bernabeumarikoneu}). \footnote{Eqn. (\ref{bernabeumekagoentuputasombrahijoputa}) is certainly valid also beyond the limit $\hbar\to 0$. The precise meaning of the above statement is that, had one computed the general Wigner integral (\ref{bernabeumarikoneu}) only within the WKB approximation, and still under the assumptions (\ref{bernabeuketefostieu}) and (\ref{bernabeumafioseudemierdeu}), the same result (\ref{bernabeumekagoentuputasombrahijoputa}) would have been obtained. This is so because the WKB approximation to the harmonic oscillator is  actually exact.}
 
It is therefore natural to ask, under what conditions is it possible to transform any given quantum--mechanical state into the semiclassical regime? Beyond the case of Hamiltonians that are at most quadratic in $q$ and $p$, it is by no means obvious that such a transformation can be made. Our central claim is that such a transformation can {\it always}\/ be made, provided that one appropriately transforms the phase--space wavefunction into the right variables. We will prove that the property of {\it local symplectic covariance of quantum--mechanical states}, to be defined presently,  ensures the possibility of transforming any given quantum--mechanical state into the semiclassical regime; this is the gauge symmetry alluded to after eqn. (\ref{ramallochupameelcarallo}). In fact we have already established this conclusion in refs. \cite{PHASEGERBESDOS, NOI, HIGGS} using the abstract mathematics of gerbes. However in this paper we will develop the alternative, though equivalent, viewpoint sketched above: a theory of symplectic connections on phase space. For a detailed account of the symplectic viewpoint see refs. \cite{LETTER, GOSSONDOS}.  Phase--space quantum mechanics is also closely related to deformation quantisation \cite{DQ}. Related matters, not always primarily concerned with quantum mechanics (some as far afield as quantum gravity) are dealt with in refs. \cite{MATONE, COMPEAN, ERCOLESSI, MINIC}. We believe that clarifying the quantum--mechanical issues raised here may substantially contribute to such apparently disparate fields.

The reader should be warned that expressions such as {\it connections, gauge invariance, symplectic covariance}\/ and the like do {\it not}\/ refer to standard Yang--Mills gauge theory on fibre bundles, but rather to a {\it gerbe}\/ gauge theory.  This notwithstanding, in this paper we will renounce the mathematical language of gerbes in favour of the physical language of Wigner functions.

\section{Configuration--space wavefunctions  {\it vs.} phase--space states}\label{bernabeuhijoputeu}

Let a $2d$--dimensional phase space $\mathbb{P}$ be given. We can pick local Darboux coordinates $q^j, p_j$ such that the symplectic form reads \footnote{We will denote $q^j, p_j$ collectively by $q,p$, omitting all sums over the $2d$ dimensions of $\mathbb{P}$.}
\begin{equation}
\omega={\rm d}q\wedge{\rm d}p,
\label{bienn}
\end{equation}
The canonical 1--form $\theta$ on $\mathbb{P}$ defined as \cite{MS}
\begin{equation}
\theta:=-p{\rm d}q
\label{afzct}
\end{equation}
satisfies 
\begin{equation}
{\rm d}\theta=\omega.
\label{tgkmjnmm}
\end{equation}
We will also need the integral invariant of Poincar\'e--Cartan, denoted $\lambda$. If $H$ denotes the Hamiltonian function, then $\lambda$ is 
defined as \cite{MS}
\begin{equation}
\lambda:=\theta+H{\rm d}t.
\label{komome}
\end{equation}
The mechanical action equals (minus) the line integral of $\lambda$,
\begin{equation}
S=-\int\lambda.
\label{swws}
\end{equation}
On constant--energy submanifolds of $\mathbb{P}$, or else for fixed values of the time, we have
\begin{equation}
{\rm d}\lambda=\omega, \qquad H={\rm const.}
\label{kuadraos}
\end{equation}
By eqn. (\ref{swws}) we can perform the transformation
\begin{equation}
\lambda\longrightarrow\lambda+{\rm d}f,\qquad f\in C^{\infty}(\mathbb{P}),
\label{bbmj}
\end{equation}
where $f$ is an arbitrary function on $\mathbb{P}$ with the dimensions of an action, without altering the classical mechanics defined by $\omega$. The transformation (\ref{bbmj}) amounts to shifting $S$ by a constant $C$,
\begin{equation}
S\longrightarrow S+C,\qquad C:=-\int{\rm d}f.
\label{chif}
\end{equation}
The way the transformation (\ref{bbmj}) acts on the quantum theory is well known. In the WKB approximation, the wavefunction reads \cite{LANDAUQM}
\begin{equation}
\psi_{\rm WKB}=R\exp\left(\frac{\rm i}{\hbar}S\right)
\label{ktfyrmlldmrd}
\end{equation}
for some amplitude $R$. Thus the transformation (\ref{bbmj}) multiplies the WKB wavefunction $\psi _{\rm WKB}$ and, more generally, any wavefunction $\psi$, by the {\it constant}\/ phase factor $\exp\left({\rm i}{C}/{\hbar}\right)$:
\begin{equation}
\psi\longrightarrow \exp\left(\frac{{\rm i}}{{\hbar}}{C}\right)\psi.
\label{llkbkb}
\end{equation}
Gauging the rigid symmetry (\ref{llkbkb}) one obtains the transformation law 
\begin{equation}
\psi(q)\longrightarrow\Psi_f(q,p):= \exp\left(-\frac{{\rm i}}{{\hbar}}f(q,p)\right)\psi(q), \qquad f\in C^{\infty}(\mathbb{P}),
\label{llmerk}
\end{equation}
$f$ being an arbitrary function on phase space, with the dimensions of an action. We will henceforth  call the objects $\Psi_f(q,p)$ {\it phase--space states}.  The Born interpretation of the wavefunction is maintained since $\vert\Psi_f\vert^2=\vert\psi\vert^2$, {\it i.e.}, the probability density remains unchanged.\footnote{The normalisation integral of  $\Psi_f(q,p)$ is to be understood as $\int{\rm d}p{\rm d}q\vert\Psi_f(q,p)\vert^2=V\int{\rm d}q\vert\psi(q)\vert^2$, where $V$ is the (possibly infinite) volume of the momentum subspace of phase space, which one cancels.} However the action of any given operator $\Omega(q,p)$ on $\Psi_f(q,p)$ will in general differ from the action of the same $\Omega(q,p)$ on $\psi(q)$. We also remark that an arbitrary function $\Phi(q,p)$ on phase space does {\it not}\/ qualify as a state unless it can be factorised as in eqn. (\ref{llmerk}), with $\psi(q)$ square--integrable.

Now eqn. (\ref{llmerk}) implies that, if the original wavefunction $\psi$ depends only on the coordinates $q$, its transform $\Psi_f$ under an arbitrary $f\in C^{\infty}(\mathbb{P})$ generally depends also on the momenta $p$. The question arises,  are the transformations (\ref{llmerk}) a symmetry of the theory? That is, are we allowed to perform the gauging (\ref{llmerk})? In section \ref{lacaspaloscojones} we will answer this question in the affirmative. In the meantime we observe that eqn. (\ref{llmerk}) allows us to arbitrarily pick, on a point--by--point basis on phase space, the zero point for the classical action $S$. The symmetry (\ref{llmerk}) is reminiscent of the U(1) gauge symmetry of electromagnetism. However there need be no electric charge present in our setup. Moreover, while electromagnetism is the gauge theory of a U(1) fibre bundle over spacetime, the gauge theory under consideration here is not of the Yang--Mills type, as will become evident presently. The gauge transformation (\ref{llmerk}) is definitely {\it not}\/ that of electromagnetic theory.

\section{The Schroedinger equation on phase space}\label{lacaspaloscojones}

It has been shown in ref. \cite{LETTER} that  the Schr\"odinger equation for the wavefunction $\psi(q)$ on configuration space,
\begin{equation}
H\left(q,-{\rm i}\hbar{\partial_q}\right)\psi(q)=E\psi(q),
\label{neu}
\end{equation}
is equivalent  to the following Schr\"odinger--like equation for the state $\Psi_f(q,p)$ on phase space \cite{VEGA}:
\begin{equation}
H\left(\frac{q}{2}+{\rm i}\hbar{\partial_p}, \frac{p}{2}-{\rm i}\hbar{\partial_q}\right)\Psi_f(q,p)=E\Psi_f(q,p).
\label{lvzmsrb}
\end{equation}
The operators within the argument of the Hamiltonian (\ref{lvzmsrb}) are those already encountered in eqn. (\ref{losdebernabeusoisunapanda}),
so eqn. (\ref{lvzmsrb}) can be rewritten as
\begin{equation}
H\left(Q_{A_0'}, P_{A_0'}\right)\Psi_f(q,p)=E\Psi_f(q,p).
\label{vzzhjp}
\end{equation}
We have also observed that $Q_{A_0'}$,  $P_{A_0'}$ satisfy the Heisenberg algebra (\ref{bernabeumekagueuentuputamadreu}). A computation shows that $\Psi_f(q,p)$ in (\ref{lvzmsrb}) and $\psi(q)$ in (\ref{neu}) are related as per eqn. (\ref{llmerk}), the argument $f(q,p)$ of this latter exponential being
\begin{equation}
f_{A_0'}(q,p):=\frac{1}{2}pq.
\label{maurix}
\end{equation}
That is, the Schr\"odinger eqns. (\ref{neu}) and (\ref{lvzmsrb}) are equivalent, and the respective state $\Psi_f(q,p)$ and wavefunction $\psi(q)$ are related as
\begin{equation} 
\Psi_f(q,p)=\exp\left(-\frac{{\rm i}}{2{\hbar}}pq\right)\psi(q).
\label{laer}
\end{equation}
Moreover, the connection $A_0'$ of eqn. (\ref{ketefolleubernabeu}) is in fact the differential of the function (\ref{maurix}):
\begin{equation}
A_0'=\frac{1}{{\rm i}\hbar}{\rm d}f_{A_0'}=\frac{1}{2{\rm i}\hbar}\left(p\,{\rm d}q+q\,{\rm d}p\right).
\label{goss}
\end{equation}
Covariantising the symplectic derivative (\ref{chupapollasdemierda}) with the connection (\ref{ketefolleubernabeu}) is equivalent to the symplectic transformation considered in ref. \cite{LETTER} that renders the quantum theory manifestly symmetric under the symplectic exchange of $q$ and $p$. This latter symmetry is conspicuously absent in the usual formulation of quantum mechanics based on the Schr\"odinger equation (\ref{neu}) on configuration space.  

As discussed in section \ref{bernabeudemierdeu}, one can consider more general covariantisations of the symplectic derivative ${\rm d}'$. Given a solution $\psi=\psi(q)$ of the Schr\"odinger equation (\ref{neu}) on configuration space, and given a function $f_{A'}\in C^{\infty}(\mathbb{P})$, define $\Psi_{f_{A'}}(q,p)$ as per eqn. (\ref{llmerk}). We can require the latter to satisfy a phase--space Schr\"odinger equation, that we can determine as follows. One picks a certain connection (\ref{ramallochupameelcarallo}) and constructs the covariant symplectic derivative (\ref{bernabeuchupameelboleu}). The components $A_q'=A'_q(q,p)$ and $A_p'=A'_p(q,p)$ are unknown functions of $q,p$. However they are not totally unconstrained, because the position and momentum operators 
\begin{equation}
Q_{A'}:=A_p'+{\rm i}\hbar\partial_p, \qquad P_{A'}:=A_q'-{\rm i}\hbar\partial_q
\label{edfre}
\end{equation}
will enter the Hamiltonian $H(Q_{A'}, P_{A'})$ obtained from $H(Q=q, P=-{\rm i}\hbar\partial_q)$ by the replacements $Q\rightarrow Q_{A'}$, $P\rightarrow P_{A'}$: 
\begin{equation}
H\left(Q_{A'}, P_{A'}\right)=\frac{1}{2m}P_{A'}^2+V(Q_{A'})=\frac{1}{2m}\left(A_q'-{\rm i}\hbar\partial_q\right)^2+V(A_p'+{\rm i}\hbar\partial_p).
\label{tse}
\end{equation}
As such, the operators (\ref{edfre}) must satisfy the canonical commutation relations (\ref{bernabeumekagueuentuputamadreu}). This requires that the following {\it integrability condition}\/ hold:
\begin{equation}
\frac{\partial A'_{p_j}}{\partial q^l}+\frac{\partial A'_{q^l}}{\partial p_j}=\delta^j_l.
\label{frz}
\end{equation}
Notice the positive sign, instead of negative, between the two summands on the left--hand side of eqn. (\ref{frz}). This is ultimately due to the fact that we are covariantising the symplectic derivative ${\rm d}'$ rather than the usual exterior derivative ${\rm d}$. For this reason, contrary to what one would expect in Yang--Mills theory, the integrability condition (\ref{frz}) is {\it not}\/ a constant--curvature condition. This is a consequence of the fact, already mentioned, that the gauge symmetry at hand is not of the Yang--Mills type. Now a computation shows that the phase--space Schr\"odinger equation
\begin{equation}
H(Q_{A'}, P_{A'})\Psi_f(q,p)=E\Psi_f(q,p)
\label{slke}
\end{equation}
is equivalent to the Schr\"odinger equation (\ref{neu}) on configuration space if, and only if, $A_q'$, $A_p'$ and $f_{A'}$ are related as
\begin{equation}
A_q'=\partial_q f_{A'},\qquad A_p'=q-\partial_pf_{A'}.
\label{ttraex}
\end{equation}
Whenever eqn. (\ref{ttraex}) holds,  the integrability condition (\ref{frz}) is automatically satisfied. We conclude that picking one function $f_{A'}\in C^{\infty}(\mathbb{P})$ and defining the connection $A'$ as per eqns. (\ref{ramallochupameelcarallo}), (\ref{ttraex}), we arrive at the phase--space wave equation (\ref{slke}). Alternatively, given a connection (\ref{ramallochupameelcarallo}) and a phase--space wave equation (\ref{slke}), we can find a function $f_{A'}\in C^{\infty}(\mathbb{P})$, defined by (\ref{ttraex}) up to integration constants, such that the corresponding phase--space state $\Psi_f(q,p)$ is related to the wavefunction $\psi(q)$ as per eqn. (\ref{llmerk}), where $f=f_{A'}$. Eqn. (\ref{ttraex}) above gives us a whole $C^{\infty}(\mathbb{P})$'s worth of phase--space Schr\"odinger equations, one per each choice of a function $f_{A'}$. The latter may well be termed the {\it generating function}\/ for the transformation  (\ref{llmerk}) between configuration--space and phase--space states and their corresponding Schr\"odinger equations.

To summarise, {\it gauging the rigid symmetry (\ref{llkbkb}), {\it i.e.}, allowing for the local transformations (\ref{llmerk}), one arrives naturally at a phase--space formulation of quantum mechanics}. This answers the question, posed after at the end of section \ref{bernabeuhijoputeu}, in the affirmative:  under the assumption (\ref{llmerk}), that  symplectic covariance is a symmetry of our theory,  we arrive at the same phase--space Schroedinger equation of refs. \cite{LETTER, GOSSONDOS, VEGA}. We may therefore take symplectic covariance as our starting point.

\section{Action of the U(1) symmetry on Wigner functions}\label{mekagoentuputokaretodemierdabernabeu}

{}Further insight into the meaning of the local U(1) rotations (\ref{llmerk}) can be gained from the following observation.  The integrand of the Wigner function (\ref{bernabeumarikoneu}) factorises as the product of $\psi(q+y/2)\exp(-{\rm i}py/2\hbar)$ and $\psi^*(q-y/2)\exp(-{\rm i}py/2\hbar)$. Now
$$
\exp\left(-\frac{{\rm i}}{2\hbar}py\right)\psi\left(q+\frac{y}{2}\right)=\exp\left(-\frac{{\rm i}}{2\hbar}py\right)\exp\left(q\frac{\partial}{\partial(y/2)}\right)\psi\left(\frac{y}{2}\right)
$$
$$
=\exp\left(\frac{{\rm i}}{\hbar}qp\right)\exp\left(q\frac{\partial}{\partial (y/2)}\right)\exp\left(-\frac{{\rm i}}{2\hbar}py\right)\psi\left(\frac{y}{2}\right)
$$
$$
=\exp\left(\frac{{\rm i}}{\hbar}qp+q\frac{\partial}{\partial (y/2)}\right)\exp\left(-\frac{{\rm i}}{2\hbar}py\right)\psi\left(\frac{y}{2}\right)
$$
\begin{equation}
=\exp\left[\frac{{\rm i}}{\hbar}q\left(p+\frac{\hbar}{{\rm i}}\frac{\partial}{\partial (y/2)}\right)\right]\Psi_{g}\left(\frac{y}{2},p\right),
\label{pepebernabeuketefostieu}
\end{equation}
where the Baker--Campbell--Hausdorff formula has been used, and
\begin{equation}
\Psi_{g}\left(\frac{y}{2},p\right):=\exp\left(-\frac{{\rm i}}{\hbar}g\left(\frac{y}{2},p\right)\right)\psi\left(\frac{y}{2}\right),\qquad g(y,p):=yp.
\label{bernabeuramallo,ramallobernabeu:dosmarikonesdemierdeu}
\end{equation}
Recalling eqn. (\ref{losdebernabeusoisunapanda}), we can rewrite (\ref{pepebernabeuketefostieu}) as
$$
\exp\left(-\frac{{\rm i}}{2\hbar}py\right)\psi\left(q+\frac{y}{2}\right)=\exp\left[\frac{2{\rm i}q}{\hbar}\left(\frac{p}{2}-{\rm i}\hbar\frac{\partial}{\partial y}\right)\right]\Psi_{g}\left(\frac{y}{2},p\right)
$$
\begin{equation}
=\exp\left[\frac{2{\rm i}q}{\hbar}P_{A'_0}(y,p)\right]\Psi_{g}\left(\frac{y}{2},p\right),
\label{barbonchupameelbolon}
\end{equation}
where the arguments $y,p$ within $P_{A'_0}(y,p)$ remind us that $y$ replaces $q$ as the variable being differentiated. By the same token,
\begin{equation}
\exp\left(-\frac{{\rm i}}{2\hbar}py\right)\psi^*\left(q-\frac{y}{2}\right)=\exp\left[-\frac{2{\rm i}q}{\hbar}P_{A'_0}(y,p)\right]\Psi_{g}^*\left(-\frac{y}{2},p\right).
\label{barbonkabron}
\end{equation}
Altogether, eqns. (\ref{barbonchupameelbolon}) and (\ref{barbonkabron}) allow one to recast the Wigner function (\ref{bernabeumarikoneu}) as \footnote{In eqn.  (\ref{bernabeumarikoneu}) we had $d=1$.}
\begin{equation}
(2\pi\hbar)^d W_{\psi}(q,p)
\label{folleuabernabeu}
\end{equation}
$$
=\int{\rm d}y\,\Psi_{g}^*\left(-\frac{y}{2},p\right)\exp\left[-\frac{2{\rm i}q}{\hbar}\overleftarrow P_{A'_0}(y,p)\right]\exp\left[\frac{2{\rm i}q}{\hbar} \overrightarrow P_{A'_0}(y,p)\right]\Psi_{g}\left(\frac{y}{2},p\right),
$$
where the covariant derivatives (\ref{losdebernabeusoisunapanda}) and the phase--space wavefunctions (\ref{llmerk}) appear explicitly. The arrows above $P_{A_0'}$ indicate left or right action. Also, the function $g$ of (\ref{bernabeuramallo,ramallobernabeu:dosmarikonesdemierdeu}) is twice the function $f_{A_0}'$ of (\ref{maurix}). In Dirac's notation we can reexpress (\ref{folleuabernabeu}) as
\begin{equation}
(2\pi\hbar)^d W_{\psi}(q,p)
\label{barbonmarikonazodemierda}
\end{equation}
$$
=\langle\Psi_{2f_{A_0'}}\left(-\frac{y}{2},p\right)\exp\left(-\frac{2{\rm i}q}{\hbar} P_{A'_0}(y,p)\right)\vert\exp\left(\frac{2{\rm i}q}{\hbar} P_{A'_0}(y,p)\right)\Psi_{2f_{A_0'}}\left(\frac{y}{2},p\right)\rangle.
$$
We stress that  (\ref{barbonmarikonazodemierda}) is just a symbolic rewriting of the Wigner function (\ref{folleuabernabeu}). That $W_{\psi}(q,p)$ is {\it not}\/ positive definite is reflected in the fact that (\ref{barbonmarikonazodemierda}) is {\it not}\/ the norm squared of the ket vector ({\it Wigner ket})
\begin{equation}
\vert\exp\left(\frac{2{\rm i}q}{\hbar} P_{A'_0}(y,p)\right)\Psi_{2f_{A_0'}}\left(\frac{y}{2},p\right)\rangle,
\label{wigket}
\end{equation}
since the argument of $\Psi$ in the corresponding bra vector in (\ref{barbonmarikonazodemierda}) is evaluated at $-y/2$ instead of $y/2$. However, eqn. (\ref{barbonmarikonazodemierda}) bears out very explicitly the fact, already explained in section \ref{bernabeudemierdeu}, that Wigner functions carry a symplectic connection  hidden inside. 

We can now write down the most general Wigner ket (and, with it, the most general Wigner function, eqn. (\ref{ramallokabronketefollen}) below):
\begin{equation}
\vert\exp\left(\frac{2{\rm i}q}{\hbar} P_{A'}(y,p)\right)\Psi_{f_{A'}}\left(\frac{y}{2},p\right)\rangle.
\label{ketwig}
\end{equation}
The elements entering (\ref{ketwig}) are the following.  Pick a function $f_{A'}(q,p)$ on phase space and a configuration--space wavefunction $\psi(q)$. Transform the latter into the phase--space wavefunction
\begin{equation}
\Psi_{f_{A'}}\left(\frac{y}{2},p\right):=\exp\left(-\frac{{\rm i}}{\hbar}f_{A'}(y,p)\right)\psi\left(\frac{y}{2}\right).
\label{fesgte}
\end{equation}
Next use eqn. (\ref{ttraex}) to construct the connection $A'$ corresponding to the function $f_{A'}$, and use it to covariantise the symplectic derivative. This gives the covariant position and momentum operators of eqn. (\ref{edfre}). Finally exponentiate the covariant momentum $P_{A'}$ and act with it on the wavefunction (\ref{fesgte}). The result is the Wigner ket (\ref{ketwig}), and the scalar product with its corresponding Wigner bra (after replacing $y/2\rightarrow -y/2$) is the Wigner function
\begin{equation}
(2\pi\hbar)^d W_{\Psi_{f_{A'}}}
\label{ramallokabronketefollen}
\end{equation}
$$
=\langle\Psi_{f_{A'}}\left(-\frac{y}{2},p\right)\exp\left(-\frac{2{\rm i}q}{\hbar} P_{A'}(y,p)\right)\vert\exp\left(\frac{2{\rm i}q}{\hbar} P_{A'}(y,p)\right)\Psi_{f_{A'}}\left(\frac{y}{2},p\right)\rangle
$$
$$
=\int{\rm d}y\,\Psi_{f_{A'}}^*\left(-\frac{y}{2},p\right)\exp\left[-\frac{2{\rm i}q}{\hbar}\overleftarrow P_{A'}(y,p)\right]\exp\left[\frac{2{\rm i}q}{\hbar} \overrightarrow P_{A'}(y,p)\right]\Psi_{f_{A'}}\left(\frac{y}{2},p\right).
$$

\section{Gauge transformations by 0--forms and by 1--forms}\label{pepebernabeuveteatomarporculeu}

We have so far considered the following transformations of the canonical 1--form $\theta$: 
\begin{equation}
\theta\longrightarrow\theta+{\rm d}f, \qquad f\in C^{\infty}(\mathbb{P}).
\label{bernabeutepartireelputocareu}
\end{equation}
Once integrated, these transformations gave rise to eqn. (\ref{llkbkb}); gauging the latter led to eqn. (\ref{llmerk}) and the ensuing construction. The gauge parameter of these transformations is an arbitrary function $f$, or 0--form, on phase space. 

Consider gauge--transforming the canonical 1--form $\theta$ as per
\begin{equation}
\theta\longrightarrow\theta+\xi, \quad \xi\in \Omega^1(\mathbb{P}),\quad {\rm d}\xi=0,
\label{pepebernabeumafioseudemierdeu}
\end{equation}
$\xi$ being an arbitrary closed 1--form with the dimensions of an action. \footnote{The transformations (\ref{bernabeutepartireelputocareu}) and 
(\ref{pepebernabeumafioseudemierdeu}) were respectively called $\delta_0$ and $\delta_1$ gauge transformations in ref. \cite{NOI}.} By eqn. (\ref{tgkmjnmm}), the above does not modify the dynamics defined by the symplectic form $\omega$ on phase space $\mathbb{P}$. Now, in general, $\xi$ need not be a total derivative ${\rm d}f$. Hence the line integral $\int\xi$ depends not only on the endpoints but also on the actual path taken between those two endpoints. On the other hand, the (reduced, {\it i.e.}, time--independent) mechanical action $S$ equals $-\int\theta$, the integral depending also on the path taken. Hence (\ref{pepebernabeumafioseudemierdeu}) is the most general gauge transformation possible for the canonical 1--form $\theta$: a path--dependent gauge transformation. Only when $\xi$ is a total derivative, $\xi={\rm d}f$, is this gauge transformation path--independent. Gauge transformations by 1--forms thus include gauge transformations by 0--forms, and are therefore more general.

So not only do we gauge--transform the phase of the wavefunction (gauge transformations by 0--forms); we also gauge--transform the canonical 1--form $\theta$ (gauge transformations by 1--forms). We have so far understood our construction as the gauge theory of U(1) phase transformations of the wavefunction, eqn. (\ref{llmerk}). As seen in this section, a deeper understanding is gained by regarding our construction as the theory of gauge transformations of the canonical 1--form $\theta$, by arbitrary closed 1--forms $\xi$. Such gauge transformations preserve the symplectic form $\omega$ because ${\rm d}\xi=0$. However they are not to be confused with symplectomorphisms of $\mathbb{P}$. 

Our gauge transformations are also not to be confused with gauge transformations in the sense of Yang--Mills theory (a potential 1--form $A$ and a field--strength 2--form $F$). A Yang--Mills gauge transformation would be $A\rightarrow A+g^{-1}{\rm d}g$, with $g$ a gauge--group valued function on $\mathbb{P}$. Our gauge group is U(1). A first guess would be to identify the Yang--Mills potential 1--form $A$ with the canonical 1--form $\theta$ and the field strength 2--form $F={\rm d}A$ with the symplectic form $\omega$. Setting $g={\rm e}^{{\rm i}\alpha}$, the Yang--Mills gauge transformation would be $A\rightarrow A+{\rm i}{\rm d}\alpha$. The term ${\rm i}{\rm d}\alpha$ is exact; at best it could be identified with the term ${\rm d}f$ in our gauge transformation (\ref{bernabeutepartireelputocareu}). Even so, the Yang--Mills gauge transformation law cannot reproduce our transformations by 1--forms, eqn. (\ref{pepebernabeumafioseudemierdeu}).

To summarise, the data we have on $\mathbb{P}$ do {\it not}\/ define a fibre bundle, and the gauge theory at hand is not of the Yang--Mills type. Instead, as shown in refs. \cite{PHASEGERBESDOS, NOI, HIGGS}, our gauge theory is of the gerbe type . Moreover, this is imposed on us by the fact that the natural connection sitting inside the Wigner function is symplectic. This fact suggests defining the following transformation law for $A'$ under eqn. (\ref{bernabeutepartireelputocareu}): 
\begin{equation}
A'\longrightarrow A'+{\rm d}'f, \qquad f\in C^{\infty}(\mathbb{P}).
\label{ramalloduchatekehuelesamierda}
\end{equation}
The right--hand side of (\ref{ramalloduchatekehuelesamierda}) contains the symplectic derivative ${\rm d}'$ instead of the exterior derivative d. This makes sense from what has been said so far, but we can further support this fact if we observe that the gauge--transformed symplectic connection $A'$ should continue to satisfy the integrability condition (\ref{frz}). The latter would no longer be satisfied if the right--hand side of  (\ref{ramalloduchatekehuelesamierda}) contained the usual exterior derivative.

\section{Discussion}\label{bernabonmarikon}

It has been known for long \cite{WIGNER} that Wigner functions provide a formulation of quantum mechanics that resembles classical statistical mechanics. In this paper we have established a correspondence between Wigner functions, on the one hand, and the phase--space Schroedinger equation, on the other. This correspondence is expressed by eqn. (\ref{ramallokabronketefollen}).

The phase--space Schroedinger equation actually corresponds to the choice of an irreducible, unitary representation of the Heisenberg algebra \cite{LETTER}.  This representation is given explicitly in eqn. (\ref{losdebernabeusoisunapanda}) and its generalisation (\ref{edfre}). By the Stone--von Neumann theorem, this latter representation is unitarily equivalent to the usual one $Q\psi(q)=q\psi(q)$, $P\psi(q)=-{\rm i}\hbar\partial_q\psi(q)$. This notwithstanding, we find representation (\ref{losdebernabeusoisunapanda}) and its generalisation (\ref{edfre}) more useful for our purposes. Moreover, we have seen in section \ref{bernabeudemierdeu} that the representations (\ref{losdebernabeusoisunapanda}) and (\ref{edfre}) have a natural origin in the Wigner functions (\ref{bernabeumarikoneu}) and (\ref{ramallokabronketefollen}) respectively.

Symplectic covariance is the symmetry that allows one to gauge--transform the canonical 1--form $\theta$ (\ref{afzct}) by arbitrary closed 1--forms $\xi$, as in eqn. (\ref{pepebernabeumafioseudemierdeu}). Let $\Omega^1(\mathbb{P})$ denote the set of all closed 1--forms $\xi$ on phase space $\mathbb{P}$. Then $\Omega^1(\mathbb{P})$ is the parameter space for the gauge transformations considered here. The symplectic form $\omega$ is preserved under all gauge transformations of $\theta$ by elements of $\Omega^1(\mathbb{P})$. Since 1--forms and symplectomorphisms are different objects, the theory under consideration here is different from the classical theory of canonical transformations on phase space. We have also proved that our gauge transformations are not those of Yang--Mills theory.

We have called the above transformations {\it gauge}\/ because the 1--forms $\xi\in\Omega^1(\mathbb{P})$ are obviously point--dependent. Now the (reduced, {\it i.e.}, time--independent) classical action $S$ equals the line integral $-\int\theta$, see (\ref{komome}), (\ref{swws}). Also, through Feynman's path integral, quantum--mechanical amplitudes depend not on $\theta$ but on its line integral $S$. Therefore the {\it gauge}\/ property disappears both clasically and quantum--mechanically.

One may decide to restore the missing point--dependence of these transformations, as in eqn. (\ref{llmerk}).\footnote{The fact that this symmetry is spontaneously broken in Nature \cite{HIGGS} is immaterial to the present discussion.} Actually this gauge property is called for, {\it e.g.}, in the passage from the configuration--space Schroedinger equation (\ref{neu}) to its phase--space counterpart (\ref{lvzmsrb}). Now the latter has been derived on general grounds \cite{VEGA},   independently of symplectic covariance. Moreover, eqn. (\ref{neu}) and its counterpart (\ref{lvzmsrb}) have been shown to be equivalent \cite{LETTER}. Therefore it is legal to restore the missing point--dependence of these transformations, and to take the symmetry expressed by eqn. (\ref{llmerk}) as our starting point. We call the symmetry (\ref{llmerk}) {\it local symplectic covariance of quantum--mechanical states}. Thus gauging the transformations of (the line--integrated) $\theta$ by elements of $\Omega^1(\mathbb{P})$ one arrives at the possibility to U(1)--rotate the configuration--space wavefunction $\psi(q)$ into the phase--space wavefunction $\Psi(q,p)$, as in eqn. (\ref{llmerk}). This rotation carries a point--dependent rotation parameter.

We have thus shown that local symplectic covariance of quantum--mechanical states makes it possible to U(1)--rotate the Schr\"odinger equation from configuration space into phase space, and also within phase space itself, with a point--dependent rotation parameter. We have also exhibited the presence of symplectic connections within the Wigner function. The resulting phase--space formulation of quantum mechanics looks somewhat clumsy at first, if only notationally.  And there is no getting around the Stone--von Neumann theorem. It is legitimate to ask,  what is the payoff? 

The payoff is the possibility of transforming any given quantum--mechanical state into the semiclassical regime, in which $\hbar\to 0$, after an appropriate choice of variables. The foregoing statement looks shocking on first sight. It is actually a corollary to a farther--reaching statement, to the effect that the notion of an elementary quantum is not universal but rather depends on the observer (see \cite{MINIC} and refs. therein).\footnote{The terms {\it coordinates}, {\it observer} and the like are not necessarily used here with the same meaning as in general relativity. Thus, {\it e.g.}, the term {\it observer}\/ may refer to the choice of a complex structure on phase space (whenever possible), or to a choice of gauge $\xi$, etc; see \cite{SONG} for this viewpoint.}  
In order to justify our answer let us observe that the transformations (\ref{pepebernabeumafioseudemierdeu}), and their gauged version after taking the corresponding line integral, allow us to arbitrarily pick the origin for the mechanical action $S$ on a point--by--point basis on phase space. On the other hand, large values of the quotient $S/\hbar$ correspond to the semiclassical regime, while small values of $S/\hbar$ correspond to the strong--quantum regime. Clearly these two regimes are locally interchangeable by virtue of the transformations (\ref{pepebernabeumafioseudemierdeu}); see refs. \cite{PHASEGERBESDOS, NOI, HIGGS} for more details. This same conclusion can also be arrived at differently in our approach via symplectic connections and Wigner functions. Namely, the symplectic connection (\ref{ketefolleubernabeu}) within the Wigner function (\ref{bernabeumarikoneu}) is the semiclassical limit of the general symplectic connection (\ref{ttraex}) within the corresponding Wigner function (\ref{ramallokabronketefollen}). Now, using the transformation law (\ref{ramalloduchatekehuelesamierda}) one can readily prove that the symplectic connection (\ref{ketefolleubernabeu}) is gauge equivalent to the general symplectic connection (\ref{ttraex}) under 0--form  gauge transformations. This gauge equivalence under 0--forms generally holds only locally on phase space; in the absence of homological obstructions, it will also hold globally. Hence our statement concerning the gauge equivalence of the semiclassical and the strong quantum regimes follows. 

Thus our main result is a presentation of quantum mechanics in which the statement above, to the effect that it is always possible to locally transform any state into the semiclassical regime, is explicitly realised. In the optical analogy used in section \ref{bernabeudemierdeu}, this amounts to the possibility of locally transforming into a set of variables in which Fresnel's optics (the analogue of the WKB approximation in mechanics) is a sufficiently good description of observed phenomena. While certainly meaningless within the realm of optics, these notions do have a meaning in any quantum theory of gravity \cite{QG}. In plain words, one may think that relativising the notion of a quantum (as done here) is dual to quantising gravity. In this sense, here we have developed the quantum mechanics that is pertinent to a theory of quantum gravity.

To summarise, there is interesting geometry and physics underlying quantum mechanics on phase space, some of which has been unravelled here.

\vskip1cm
\noindent {\bf Acknowledgements}  It is a great pleasure to thank Max--Planck--Institut f\"ur Gravitationsphysik, Albert--Einstein--Institut (Golm, Germany) for hospitality during the preparation of this article. This work has been supported by Ministerio de Educaci\'on y Ciencia through grant FIS2005--02761 and by Generalitat Valenciana (Spain).

\end{document}